\documentclass{mn2e}
\usepackage{natbib}
\usepackage{amsfonts}
\usepackage{amsmath}
\usepackage{amssymb}
\usepackage{graphicx}
\usepackage{color}

\newcommand{\be}{\begin{equation}}
\newcommand{\ee}{\end{equation}}

\title[Two-dimensional excursion set]
      {Stochastic bias in multi-dimensional excursion set approaches} 

\author[E. Castorina \& R. K. Sheth]
{Emanuele Castorina$^{1}$\thanks{E-mail: ecastori@sissa.it} \& Ravi K. Sheth$^{2,3}$\\  
 $^1$ SISSA - International School For Advanced Studies,
      Via Bonomea, 265 34136 Trieste, Italy\\
 $^2$ The Abdus Salam International Center for Theoretical Physics,
      Strada Costiera, 11, Trieste 34151, Italy\\
 $^3$ Center for Particle Cosmology, University of Pennsylvania, 
      209 S. 33rd St., Philadelphia, PA 19104, USA}
\begin{document}
\pagerange{\pageref{firstpage}--\pageref{lastpage}}

\maketitle 

\label{firstpage}

\begin{abstract}
We describe a simple fully analytic model of the excursion set 
approach associated with two Gaussian random walks:  the first 
walk represents the initial overdensity around a protohalo, and 
the second is a crude way of allowing for other factors which might 
influence halo formation.  This model is richer than that based on 
a single walk, because it yields a distribution of heights at first 
crossing.  We provide explicit expressions for the unconditional first 
crossing distribution which is usually used to model the halo mass 
function, the progenitor distributions from which merger rates are 
usually estimated, and the conditional distributions from which 
correlations with environment are usually estimated.  These latter 
exhibit perhaps the simplest form of what is often called nonlocal 
bias, and which we prefer to call stochastic bias, since the new bias 
effects arise from `hidden-variables' other than density, but these 
may still be defined locally.  We provide explicit expressions for 
these new bias factors.  
We also provide formulae for the distribution of heights at first 
crossing in the unconditional and conditional cases.  In contrast 
to the first crossing distribution, these are exact, even for moving 
barriers, and for walks with correlated steps.  The conditional 
distributions yield predictions for the distribution of halo 
concentrations at fixed mass and formation redshift.  They also 
exhibit assembly bias like effects, even when the steps in the walks 
themselves are uncorrelated.  Our formulae show that without prior 
knowledge of the physical origin of the second walk, the naive estimate 
of the critical density required for halo formation which is based on 
the statistics of the first crossing distribution will be larger than 
that based on the statistical distribution of walk heights at first 
crossing; both will be biased low compared to the value associated 
with the physics.  Finally, we show how the predictions are modified 
if we add the requirement that halos form around peaks:  these depend 
on whether the peaks constraint is applied to a combination of the 
overdensity and the other variable, or to the overdensity alone.  
Our results demonstrate the power of requiring models to reproduce not 
just halo counts but the distribution of overdensities at fixed protohalo 
mass as well.  
\end{abstract}


\section{Introduction}
\label{intro}

The excursion set approach, pioneered by \cite{epstein83} and 
developed substantially by \cite{bcek91}, \cite{lc93}, \cite{mw96} 
and \cite{rks98} yields important insight into various features of 
hierarchical clustering.  Although recent work has highlighted the 
limitations of this approach \citep{ps12}, the limitations are 
primarily of a quantitative rather than qualitative nature.  

The approach combines the statistics of the initial density fluctuation 
field with the physics of spherical or triaxial collapse, to make 
predictions for the abundance of virialized objects as a function 
of time.  This means that it provides information about merger rates, 
the high-redshift progenitors of objects of fixed mass at a later time, 
the tendency for the mass function in dense regions to be top-heavy, 
and hence how the spatial clustering of these objects depends on their 
mass.  

In the spherical collapse model, the evolution of an object is 
determined by its own overdensity.  This enters in the excursion set 
approach as follows.  One associates a one-dimensional random walk 
with each position in space; this walk shows how the initial overdensity 
depends on the smoothing scale over which the density is averaged.  
The largest scale on which this walk exceeds the critical density 
required for spherical collapse contains a mass; this is the excursion 
set estimate of the mass of the object in which this particular position 
in space will end-up.  Therefore, in this approach, the technical 
problem to be solved is that of the first crossing distribution of 
a barrier whose height may depend on the number of steps taken by 
the one-dimensional random walk.  The statistics of the initial 
fluctuation field determines the ensemble of walks over which to 
average.  

In triaxial collapse models, the evolution of an object is determined 
by more than its initial overdensity \citep{bm96,smt01}.  In the context 
of such models, it is natural to ask how these extra parameters enter 
the excursion set approach.  It should come as no surprise that 
each additional variable simply adds an extra walk 
\citep{smt01, cl01, st02}, but there is no guarantee that these 
variables are Gaussian distributed.  As a result, the technical problem 
becomes one of first crossing a multi-dimensional barrier by multi-dimensional 
walks.  However, it has recently been realized that this has nontrivial, 
qualitatively different, consequences for halo bias:  in effect, the 
correlations between these other parameters on the large scale density 
field introduce what are known as nonlocal bias effects \citep{scs12}.  
In this respect, the multi-dimensional excursion set approach is 
considerably richer than the one-dimensional one.  

The main goal of this paper is to illustrate a number of these 
qualitatively new features of the multi-dimensional excursion set 
approach.  Our goal here is not so much to develop a model which 
reproduces effects seen in simulations, as to develop insight:  
therefore, the emphasis is on developing a fully analytic model in 
which it is easy to see the origin of these new effects.  It turns out 
that this model may not be that unrealistic -- this is explored further 
in \cite{arsc13}.  

Section~\ref{g2d} describes our model and provides expressions for the 
usual excursion set approach quantities, as well as for the qualitatively 
new ones.  Section~\ref{extend} describes a number of extensions, 
including an explicit calculation of how all the predictions are modified 
if protohalos are identified with peaks in the initial field.  We use 
this to demonstrate how requiring models to reproduce both halo 
counts as well as overdensities at fixed halo mass provides sharp 
constraints.  A final section summarizes.  

\section{Two independent Gaussian walks with uncorrelated steps}\label{g2d}

Let $\delta$ and $g$ both denote zero-mean Gaussian variables, 
with variance $\langle\delta^2\rangle\equiv s$ and 
$\langle g^2\rangle\equiv \beta^2 s$ respectively.  When plotted 
as a function of $s$, these represent walks associated with the 
overdensity and the second variable which matters for collapse.  
We will assume that $\delta$ and $g$ are independent: 
$\langle\delta g\rangle = 0$.   

We will use $f(s)$ to denote the distribution of $s$ when 
\begin{equation}
 \delta \ge \delta_c(s) + g
 \label{dbdc}
\end{equation}
for the first time.  We will also be interested in 
 $p(\delta_{1\!\times}|s)$, 
the distribution of walk heights at first crossing.  
The excursion set ansatz assumes that the quantity $f(s)$ is related 
to the mass fraction in halos having mass $m(s)$ by 
\begin{equation}
 f(s)\,{\rm d}s = \frac{m}{\bar\rho}\,\frac{{\rm d}n(m)}{{\rm d}m}\,{\rm d}m,
 \label{ansatz}
\end{equation}
where ${\rm d}n/{\rm d}m$ is the comoving number density of halos of 
mass $m$, and $\bar\rho$ is the comoving background density.  

\subsection{Rotation of coordinate system}
When the inequality~(\ref{dbdc}) is saturated, it defines a line in 
the $(\delta,g)$ plane.  The clearest way to think of this problem 
is to change variables to ones which run parallel and perpendicular 
to this line.  Therefore, define 
\begin{equation}
 g_- = \frac{\delta - g}{\sqrt{1+\beta^2}}\quad{\rm and}\quad
 g_+ = \frac{\beta\delta + g/\beta}{\sqrt{1+\beta^2}}.
 \label{rot}
\end{equation}
Notice that $\langle g_-^2\rangle = \langle g_+^2\rangle = s$, and 
that these variables are independent:
\begin{equation}
 \langle g_+ g_-\rangle 
 = \beta\langle \delta^2\rangle - \frac{\langle g^2\rangle}{\beta} = 0.
\end{equation}
In these variables, $g_-$ steps towards or away from the barrier, 
which has height $\delta_c(s)/\sqrt{1+\beta^2}$, 
and $g_+$ steps parallel to it.  

For what follows, it is useful to note that 
\begin{equation}
 \delta = \frac{g_- + \beta g_+}{\sqrt{1+\beta^2}}\quad{\rm and}\quad
      g = \beta \,\frac{g_+ - \beta g_-}{\sqrt{1+\beta^2}}.
\end{equation}

\subsection{Unconditional first crossing distribution}\label{universal}
The independence of $g_+$ and $g_-$ means that $f(s)$ depends only 
on $g_-$.  Since $g_-$ is just a one dimensional gaussian walk, and 
it must cross a barrier of height $\delta_c(s)/\sqrt{1+\beta^2}$, 
the first crossing distribution is that for a moving barrier, 
for which simple approximations are available \citep{st02}.  

For the special case in which $\delta_c$ does not depend on $s$, 
the first crossing distribution is 
\begin{equation}
 sf(s) = \frac{\nu f(\nu_\beta)}{2} = \nu\,\frac{\exp(-\nu^2/2)}{\sqrt{2\pi}},
 \label{sfs}
\end{equation}
where 
\begin{equation}
 \nu^2 \equiv \frac{\delta_c(0)^2/s}{1+\beta^2} \equiv \nu_\beta^2.
 \label{nubeta}
\end{equation}
Notice that $\beta=0$ yields the usual one-dimensional solution.

Notice also that the factor $1+\beta^2$ can be viewed in either of 
two ways.  Either it rescales the barrier height (which is how it 
appeared in the analysis above) or it rescales the variance $s$.  
Now, the first crossing distribution $f(s){\rm d}s$ is usually equated 
with the mass fraction in halos of mass $m$ (equation~\ref{ansatz}).  
If $\delta_c$ itself is expected to be related to the physics of halo 
formation, then the rescaling of $\delta_c$ means that one must also 
understand the physics which led to $\beta\ne 0$ if one wishes to 
derive the value of $\delta_c$ from halo abundances.  Failure to do so 
will lead to a misestimate of the true value of the value of $\delta_c$ 
which matters for the physics.  If we require $\delta_c\approx 1.686$, 
then matching halo counts requires $(1+\beta^2)^{-1} \approx 0.7$ so 
$\beta\approx 0.6$ \citep{st99}. 

\subsection{Distribution of height at first crossing}\label{d1}
Define $\delta_{1\!\times}$ to be the value of $\delta$ when 
$g_- = \delta_c(s)/\sqrt{1+\beta^2}$.  Then 
\begin{equation}
 \delta_{1\!\times} \equiv \frac{\delta_c(s)/\sqrt{1+\beta^2} + \beta g_+}
                            {\sqrt{1+\beta^2}}
     = \frac{\delta_c(s)}{1+\beta^2} + \frac{\beta\ g_+}{\sqrt{1+\beta^2}}.
 \label{d1x}
\end{equation}
Since  $g_+$ is just a Gaussian with zero mean and variance $s$ 
(recall it is independent of $g_-$), the expression above shows that 
\begin{equation}
 p(\delta_{1\!\times}|s) =
  \frac{{\rm e}^{-(\delta_{1\!\times} - \mu_{1\!\times})^2/2\Sigma_{1\!\times}^2}}
                            {\sqrt{2\pi\Sigma_{1\!\times}^2}}
 \label{pd1x}
\end{equation}
where 
\begin{equation}
 \mu_{1\!\times} = \frac{\delta_c(s)}{1+\beta^2}\quad {\rm and}\quad
 \Sigma_{1\!\times}^2 = \frac{\beta^2}{1+\beta^2}\,s.
\end{equation}
The limit $\beta=0$ yields a delta-function centered on $\delta_c(s)$ 
as it should.  

If we set $\nu_{1\!\times}\equiv \delta_{1\!\times}/\sigma$ 
where $\sigma^2\equiv s$,
and recall from equation~(\ref{nubeta}) that 
 $\nu_{\beta}\equiv (\delta_c(0)/\sigma)/\sqrt{1+\beta^2}$, 
then it is useful to think of the distribution above as 
$p(\nu_{1\!\times}|\nu_\beta)$, the conditional distribution of 
$\nu_{1\!\times}$ given $\nu_\beta$:  in this case, the expression 
above is the standard expression for the conditional Gaussian 
distribution with correlation parameter $(1+\beta^2)^{-1/2}$.

Note that equation~(\ref{d1x}), and hence equation~(\ref{pd1x}) are  
exact even when $\delta_c$ depends on $s$.  In this respect, the 
distribution of $\delta_{1\!\times}$ at first crossing is much simpler 
than is the first crossing distribution itself -- it always has a 
Gaussian shape, with the barrier only affecting the mean value of 
this Gaussian.  

It is also worth noting that
 $\langle\delta_{1\!\times}|s\rangle = \mu_{1\!\times}$ 
is {\em guaranteed} to be less than $\delta_c$.  Thus, without prior 
knowledge of the value of $\beta$, the statistical distribution of 
$\delta_{1\!\times}$ will lead to a misestimate of the value of $\delta_c$ 
which is associated with the physics.  In this context, it is useful 
to think in terms of the distribution of differences from $\delta_c$.  
If we define
 $\Delta_{1\!\times - c}\equiv \delta_{1\!\times}-\delta_c(s)$, 
then it is Gaussian distributed with mean
 $-\delta_c(s)\,\beta^2/(1+\beta^2)$ 
and variance
 $s\beta^2/(1+\beta^2)$.  I.e., the mean is $\delta_c(s)$ times the 
same factor by which $s$ is rescaled.  This provides a simple 
operational way of determining the value of $\beta$ from a 
measurement of $p(\Delta_{1\!\times -c}|s)$.  

\subsection{Distribution of the barrier at first crossing}\label{g1}
Similarly, define $g_{1\!\times}$ to be the value of $g$ at first crossing.  
Then, because $g_{1\!\times} \equiv \delta_{1\!\times} - \delta_c$, it has 
the same distribution as $\delta_{1\!\times}$, but with a shifted mean.  
Specifically, $p(g_{1\!\times})$ will be Gaussian with mean
$-\delta_c(s)\,\beta^2/(1+\beta^2)$ and variance $s\beta^2/(1+\beta^2)$.

\subsection{The two-barrier problem and progenitor distributions}\label{f10}
Symmetry means that the distribution of $S_1$ at which 
\begin{equation}
 \delta \ge \delta_{c1} + g
\end{equation}
for the first time, given that inequality~(\ref{dbdc}) was first 
satisfied on scale $S_0<S_1$, is given by equation~(\ref{sfs}) 
but with $\nu^2$ replaced by 
\begin{equation}
 \nu_{10}^2 = \frac{(\delta_{c1}-\delta_{c0})^2}{(S_1-S_0)(1+\beta^2)}.
 \label{nu10}
\end{equation}
The limit $\beta=0$ yields the usual expression for progenitor 
distributions associated with one-dimensional walks \cite[e.g.][]{lc93}.  

Halo formation is often identified with the time when at least half the 
total mass has been assembled in pieces that are each more than $\mu$ 
times the final mass.  For $\mu>1/2$, there can be only one such piece 
so the formation time distribution is given by 
\begin{equation}
 p(\delta_{cf}\ge\delta_{c1}|M,\delta_{c0})
  = \int_{\mu M}^M {\rm d}m\, \frac{M}{m}\, f(m,\delta_{c1}|M,\delta_{c0})
\end{equation}
\citep{lc93}.  For white noise initial conditions ($s\propto m^{-1}$) 
and $\mu = 1/2$ this becomes 
\begin{equation}
 p(\omega_f) = 2\omega_f\,{\rm erfc}(\omega_f/\sqrt{2})
\end{equation}
where $\omega_f = \nu_{f0}$ with $\nu_{f0}$ given by equation~(\ref{nu10}).
Because $\omega_f$ includes a factor of $1+\beta^2$, the mean formation 
redshift will be scaled to higher values than when $\beta=0$.  This 
sort of rescaling yields better agreement with measurements in 
simulations \citep{gmst07,mgs08}.  
See \cite{rks11} for the case $\mu<1/2$.  

\subsection{Conditional distributions and correlations with environment}\label{fsS}
Similarly, the distribution of $s$ at which inequality~(\ref{dbdc}) is 
first satisfied, given that $\delta$ has height $\Delta$ on some scale 
$S<s$, but $G$ is unconstrained (except by the requirement that 
$\Delta - G < \delta_{c0}$), is also given by equation~(\ref{sfs}) 
but with $\nu^2$ replaced by 
\begin{equation}
 \nu_{\Delta}^2 = \frac{(\delta_{c0}-\Delta)^2}{s(1+\beta^2) - S}.
 \label{nuD}
\end{equation}
(The Appendix provides a short derivation.)  
This can be thought of as subtracting from the variance $s(1+\beta^2)$ 
the piece which comes from constraining $\delta=\Delta$ on scale $S$, 
which makes its correspondence to the one-dimensional expression 
(the $\beta=0$ limit of this expression) obvious.  

\begin{figure}
 \centering
 \includegraphics[width=\hsize]{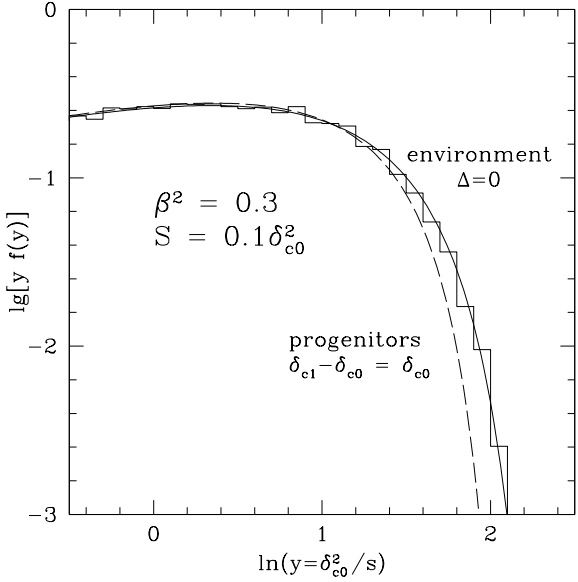}
 \caption{First crossing distributions for our two-dimensional walks 
          with $\beta^2 = 0.3$ which are conditioned to first pass through 
          $\Delta=0$ on scale $S=0.1\delta_{c0}^2$ (histogram); smooth 
          curve shows our prediction (equation~\ref{nuD} in 
          equation~\ref{sfs}).  
          The effective cosmology of this environment has critical 
          density $\delta_{c0}$; dashed curve shows the progenitor 
          distribution with this same effective cosmology 
          (equation~\ref{nu10} in equation~\ref{sfs}).  
          For one-dimensional walks, the solid curve would be the same 
          as the dashed one.
         }
 \label{evolenv}
\end{figure}

Because equation~(\ref{nu10}) is different from~(\ref{nuD}), when 
expressed as a function of $s$ rather than $\nu$, the conditional 
distribution is different from the progenitor one, whereas they 
are the same for one-dimensional walks.  The difference between 
the two is largest in the $s\to S$ limit, where the conditional 
distribution predicts more objects than does the progenitor distribution.  
Figure~\ref{evolenv} illustrates.  
self-similar distribution.  
Thus, a discrepancy between the progenitor and environmental 
dependences of clustering provides a simple way to see if stochasticity 
has played a role in determining halo abundances.  

Things are slightly more complicated if $\delta_c(s)$, of course, but 
the basic fact that progenitor and conditional distributions with 
$\delta_{c1} - \delta_{c0} = \delta_{c0}-\Delta$ will no longer be the 
same is generic.  

\subsection{Stochastic (nonlocal) bias}\label{stocbias}
The distribution of $s$ at which inequality~(\ref{dbdc}) is first 
satisfied, given that the walk was at $(\Delta,G)$ on scale $S<s$, 
is also given by equation~(\ref{sfs}) but with $\nu^2$ replaced by 
\begin{equation}
 \nu_{\Delta G}^2 = \frac{(\delta_{c0} - \Delta + G)^2}{(s-S)(1+\beta^2)}.
 \label{nuDG}
\end{equation}
This follows from the fact that the distance from a point $(x_0,y_0)$ 
to the line $ax + by + c = 0$ is  $|ax_0 + by_0 + c|/\sqrt{a^2 + b^2}$.  
Alternatively, one can view this as the same shift of origin to the 
$g_-$ walk that is made in the one-dimensional case \citep[e.g.][]{lc93}.
The expression above shows that $G$ can affect halo abundances in 
qualitatively the same way that $\Delta$ can.  

In more detail, the halo overdensity is defined by the ratio of the 
conditional expression to the unconditional one \citep{mw96}.  In our 
case, this means that 
\begin{equation}
 1 + \delta_h(\nu|\Delta,G) = \frac{\nu_{\Delta G}f(\nu_{\Delta G})}{\nu f(\nu)}.
\end{equation}
The peak-background split bias factors are the coefficients in the 
Taylor series expansion of the expression above, in the limit where 
$s\gg S$.  If we write these as 
\begin{equation}
 1 + \delta_h \equiv \sum_{i,j} B_{ij}\frac{\Delta^i}{i!} \,\frac{G^j}{j!} ,
\end{equation}
then the dependence on $G$ gives rise to what is known as nonlocal bias.  
Since $G$ may also be determined by local quantities, this is, in 
general, a misnomer.  Since it is really an effect which arises from 
the dependence of halo counts on the `hidden' stochastic variable $G$, 
we think it is more accurate to call this `stochastic' bias, which 
may or may not be local.  

Recently, \cite{mps12} have shown that cross-correlating the halo 
overdensity field with the $n$th-order Hermite polynomial 
 $H_n(\Delta/\langle\Delta^2\rangle^{1/2})$ 
is an efficient way of reconstructing the $b_n$ coefficients 
even when $\langle\Delta^2\rangle^{1/2}$ is not small.  In our case, 
cross-correlating with
 $H_i(\Delta/\langle\Delta^2\rangle^{1/2})\,H_j(G/\langle G^2\rangle^{1/2})$
yields 
\begin{equation}
 \delta_c^{i+j} B_{ij} = (-1)^j\,\nu^{i+j-1}\,H_{i+j+1}(\nu),
\end{equation}
where $\nu^2 = (\delta_{c0}^2/s)/(1+\beta^2)$. 
This reduces to the usual expression \citep{mw96,mps12} when $j=0$:   
\begin{equation}
 \delta_c^k B_{k0} \equiv \delta_c^k\,b_k = \nu^{k-1}\,H_{k+1}(\nu).
\end{equation}
Since the dependence of equation~(\ref{nuDG}) on $G$ is the same as 
that on $\Delta$, cross-correlating with $H_n(G/\langle G^2\rangle^{1/2})$ 
alone yields 
\begin{equation}
 B_{0k} \equiv c_k = (-1)^k b_k.
\end{equation}
In this respect, the stochastic (possibly nonlocal) bias model here 
is simpler than that in \cite{scs12}, where the analogue of $G$ was 
not Gaussian distributed (so the associated orthogonal polynomials 
were more complicated).  

\subsection{Assembly bias}\label{abias}
Assembly bias is the correlation between properties of protohaloes 
of fixed mass and their environment, such as those first identified 
by \cite{st04},  and studied since by many others.  While it is 
generally believed that this effect should be absent in excursion 
set models with uncorrelated steps \citep{sdmw96}, we now show that 
our two-dimensional model does exhibit assembly bias, even though 
the steps in the walks are uncorrelated.  However, we caution that 
we are not claiming that this model explains assembly bias; simply 
that assembly bias is part and parcel of the multi-dimensional 
excursion set approach, even for walks with uncorrelated steps.  

The distribution of walk heights at first crossing, given that 
$\delta=\Delta$ on scale $S$, is 
\begin{equation}
 p(\delta_{1\!\times}|s,\Delta,S)
   = \frac{{\rm e}^{-(\delta_{1\!\times}-\Delta - \mu_\Delta)^2/2\Sigma_\Delta^2}}
                     {\sqrt{2\pi\Sigma_\Delta^2}}
 \label{pd1xcond}
\end{equation}
where 
\begin{equation}
 \mu_\Delta = \frac{\delta_c(s) - \Delta}{1+\beta^2}\quad {\rm and}\quad
 \Sigma_\Delta^2 = \frac{(s-S)\beta^2}{1+\beta^2}.
\end{equation}
This is the conditional analogue of equation~(\ref{pd1x}).

\begin{figure}
 \centering
 \includegraphics[width=\hsize]{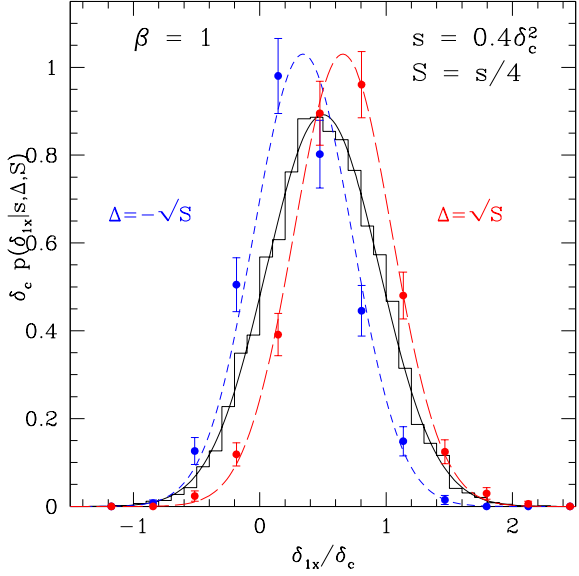}
 \caption{Dependence of walk height at first crossing, $\delta_{1\!\times}$, 
          on large scale environment.  Symbols with error bars show 
          the distribution of $\delta_{1\!\times}$ for walks which 
          first cross each other on scale $s$, and which had height 
          $\Delta$ on scale $S<s$; smooth dashed curves show 
          equation~(\ref{pd1xcond}).  Black histogram shows the 
          corresponding unconditional distribution for the same 
          value of $s$; smooth solid curve shows the corresponding 
          prediction (equation~\ref{pd1x}).
         }
 \label{cond1x}
\end{figure}

This shows that the variance is smaller than it is for 
unconditioned walks, but that the difference is negligible 
when $s\gg S$.  The mean is more interesting:  
\begin{equation}
 \langle\delta_{1\!\times}|s,\Delta,S\rangle = \Delta + \mu_\Delta
   = \frac{\delta_c(s) + \beta^2\Delta}{1+\beta^2}
\end{equation}
is shifted by $\Delta\beta^2/(1+\beta^2)$ compared to the 
unconditional mean.  Even more suggestively, this implies that 
$\langle\Delta_{1\!\times -c}|s,\Delta,S\rangle
  = [\Delta-\delta_c(s)]\,\beta^2/(1+\beta^2)$.
The dependence of this mean on the larger scale $\Delta$ is this 
model's expression of assembly bias, and is an important way in 
which the two-walk problem differs from the one-walk problem.  
When $\beta=0$ the distribution becomes a delta-function centered 
on $\delta_c$; since it is therefore independent of $\Delta$, this 
shows explicitly that the one-dimensional solution shows no assembly 
bias when the steps in the walk are uncorrelated.  

Figure~\ref{cond1x} illustrates the effect:  objects which are 
surrounded by large scale overdensities tend to have larger 
$\delta_{1x}$ than objects of the same mass in large-scale 
underdensities.  Since they have above average initial overdensities 
on scale $s$, they will also tend to have above average overdensities 
at formation (typically, on scale $\sim 2s$).  The result is a 
correlation, at fixed halo mass, between the density at formation 
and environment -- even though there will not be a correlation 
between formation time (rather than the overdensity at the formation 
time) and environment.  (In this model, as for the one-dimensional 
case, any correlation between formation time and larger scale 
environment can only come from correlations between steps.)  Since 
the density at formation is correlated with halo concentration at 
virialization \citep{nfw97}, our model predicts a correlation between 
halo concentration and environment at fixed mass.  

\section{Extensions}\label{extend}

\subsection{Correlated steps}\label{cs}
Our change of variables from $\delta,g$ to walks which step parallel 
and perpendicular to the barrier makes it straightforward to see what 
should happen when both $\delta$ and $g$ are walks with correlated  
steps.  If the correlations are the result of smoothing with the same 
filter, then the unconditional distribution $f(s)$ should be replaced 
with the corresponding expression in \cite{ms12} (see discussion 
following equation~\ref{sfspk}), but the distribution of the walk 
height at first crossing, $p(\delta_{1\!\times}|s)$, remains unchanged.  
This is because, at first crossing, $\delta_{1\!\times}$ depends only 
on $g_+$ (by definition), and $g_+$, although it has correlated steps, 
is independent of $g_-$, so it is not constrained by the fact that 
$g_- = \delta_c/(1+\beta^2)$.   

Following~\cite{pls12} the progenitor distribution should be 
well-approximated by replacing
 $\delta_{c1}-\delta_{c0} \to \delta_{c1} - (S_\times/S)\delta_{c0}$ 
and
 $(s-S)(1+\beta^2) \to [s - (S_\times/S)^2 S](1+\beta^2)$,
and the conditional distribution by replacing 
 $\delta_{c0}-\Delta \to \delta_{c0} - (S_\times/S)\Delta$ 
and 
 $s(1+\beta^2) - S \to s(1+\beta^2) - (S_\times/S)^2 S$.  
Still more accurate expressions follow from making the corresponding 
replacements in the expressions provided in~\cite{mps12}.  Testing 
these expressions is the subject of work in progress.

\subsection{Correlated Walks}\label{cw}
Suppose instead that steps in $\delta$ are uncorrelated, whereas steps 
in $g$ are correlated with those in $\delta$.  This may happen, for 
example, if the critical density for collapse depends on the overdensity 
on a larger scale, e.g. in the correlated galaxy formation model of \cite{bcfw93} 
or in theories of modified gravity \citep{lamli12}. Then, let 
\begin{equation}
 \rho \equiv \frac{\langle \delta g \rangle}
                {\langle \delta^2 \rangle ^{1/2} \; \langle g^2\rangle ^{1/2}} \;,
\end{equation}
denote the correlation parameter between $\delta$ and $g$.  
If we make the same coordinate transformation as before, then 
 $\langle g_-^2\rangle = s(1 + \beta^2 - 2\rho \beta)/(1+\beta^2)$,
 $\langle g_+^2\rangle = s(1 + \beta^2 + 2\rho \beta)/(1+\beta^2)$
and 
 $\langle g_+g_-\rangle = s\,\rho\, (1 - \beta^2)/(1+\beta^2)$.
We can always write 
 $p(g_-,g_+) = p(g_-)\,p(g_+|g_-)$, where  $p(g_+|g_-)\ne p(g_+)$ is 
a Gaussian distribution with mean
 $g_-\,\langle g_+g_-\rangle/\langle g_-^2\rangle$ 
and variance 
 $\langle g_+^2\rangle\,[1 - \langle g_+g_-\rangle^2/\langle g_-^2\rangle\langle g_+^2\rangle]$.

Since the first crossing distribution depends on $p(g_-)$ and not on $g_+$, 
it is given by the same expression as for uncorrelated walks, but with 
 $\nu^2 = (\delta_c^2/s)/(1 + \beta^2 + 2\rho\beta)$.  
This shows that the amount by which $\delta_c$ appears to be rescaled 
depends on $\beta$ as well as the correlation parameter.  
 
However, $p(\delta_{1\!\times}|s)$ will be affected.  Namely, at first 
crossing, $\delta_{1\!\times}$ is given by equation~(\ref{d1x}), so 
\begin{equation}
 \langle \delta_{1\!\times}|s\rangle = \frac{\delta_c(s)}{1+\beta^2}
  + \frac{\beta \langle g_+|g_-\rangle}{\sqrt{1+\beta^2}}
\end{equation}
where, because $g_- = \delta_c/(1+\beta^2)$ at first crossing, 
\begin{equation}
 \langle g_+|g_-\rangle
  = \frac{\delta_c(s)}{1+\beta^2}\,\rho\, 
    \frac{1 - \beta^2}{1 + \beta^2 - 2\rho\beta}\,.
\end{equation}
Therefore, $p(\delta_{1\!\times}|s)$ is Gaussian with mean and variance 
\begin{align}
  \mu\,&=\, \frac {\delta_c} {1+\beta^2} \left(1 + \rho\,
     \frac{\beta\,(1-\beta^2)}{1+\beta^2-2\beta \rho } \right) \\
  \Sigma^2\,&=\,s\,
       \frac{\beta^2\,(1 - \rho^2)}{(1+\beta^2-2\beta \rho)}. 
\end{align}
For $\rho=0$, this reduces to equation~(\ref{pd1x}); for 
$\rho = 1$ or $-1$, corresponding to complete correlation or 
anti-correlation, the distribution becomes a Dirac delta function 
centered on $\mu = 1/(1-\beta)$ or $1/(1+\beta)$, respectively.  
(This can be understood simply from the fact that, in these limiting 
cases, the two-dimensional walk is confined to a line, and this line 
can only cross the line defined by the barrier at a single point.)

It is a curious fact that when $\beta = 1$ 
(the two walks have the same variance), then there is no shift to the 
mean, and the variance becomes $s\,(1+\rho)/2$.  This can be traced 
back to the fact that, when $\beta=1$, then $\langle g_+g_-\rangle = 0$; 
i.e., the walks in $g_-$ and $g_+$ are independent (even though $\delta$ 
and $b$ are correlated), but they have different variances.  

But in general, correlations between the walks lead to a shift in the 
mean and a rescaling of the variance. However, they do not change the 
fact that $p(\delta_{1x}|\sigma)$ is Gaussian.  In practice, one should 
be able to determine if $\rho\ne 0$ because the three unknowns, 
$\delta_c$, $\beta$ and $\rho$ can be determined from our expressions 
for the mean and variance of $p(\delta_{1x}|\sigma)$ and the required 
rescaling of $s$ in the first crossing distribution $f(s)$.  

\subsection{Higher-dimensional walks and/or other distributions}\label{ng}
Our fundamental assumption, that equation~(\ref{dbdc}) accurately 
captures the physics of collapse, is, of course, only an idealization.  
Note, however, that if other variables also mattered, and they were 
also Gaussian distributed, such that equation~(\ref{dbdc}) becomes  
\begin{equation}
 \delta \ge \delta_c(s) + \sum_{i=1}^{n-1} g_i,
\end{equation}
then, because the sum of Gaussians is itself Gaussian, this $n$-dimensional 
model reduces to the 2-dimensional one we have just solved, with 
$\beta^2 = \sum_{i=1}^{n-1} \beta_i^2$.  

Alternatively, suppose instead that 
\begin{equation}
 \delta \ge \delta_c + \chi,
\end{equation} 
where $\chi$ follows a non-Gaussian distribution.  E.g., \cite{scs12}
study a model in which $\delta_c$ is independent of $s$, but $\chi^2$ 
is drawn from a chi-squared distribution with five degrees of freedom.  
However, this distribution has a mean which depends on $s$.  
If the distribution of $\Delta\chi \equiv\chi-\langle\chi\rangle$ is 
not too different from a Gaussian, then we can use our 2-dimensional 
Gaussian model as a reasonable approximation to this one, with 
$\delta_c(s)$ in equation~(\ref{dbdc}) equal to 
$\delta_c + \langle\chi\rangle$ and $g$ a zero-mean Gaussian variate 
having the same variance as $\Delta\chi$.  E.g., for the model in 
\cite{scs12},
 $\langle\chi\rangle\approx 0.95\sqrt{s}$ and
 $\langle (\Delta\chi)^2\rangle \approx 0.09s$.  
I.e., this model should be reasonably well approximated by our two-Gaussian 
model with $\delta_c(s) = \delta_c + 0.95\sqrt{s}$ and $\beta^2 = 0.09$.  

This has the following interesting consequence.  At first crossing, 
the distribution of $\Delta\chi$ will be like that of $g_{1\!\times}$, 
meaning that it should have mean and variance approximately given by 
$-\delta_c(s)\beta^2/(1+\beta^2)$ and $s\,\beta^2/(1+\beta^2)$.  
Since the variance of the initial variate $\chi$ was $\beta^2 s$, 
one should think of $\chi_{1\!\times}$ as having variance reduced by 
$(1+\beta^2)^{-1}$.  For it to still have approximately the same functional 
form as $\chi$ itself, it should have mean $0.95\sqrt{s/(1+\beta^2)}$, 
which is smaller than the original value of $0.95\sqrt{s}$.  
For $\beta\ll 1$, we can think of this as a shift in the mean by 
$-0.95\sqrt{s}\beta^2/2$.  The actual shift, 
$-\delta_c(s)\beta^2/(1+\beta^2)$, has the same sign, but a different 
amplitude, indicating that the distribution of $\chi_{1\!\times}$ will 
not be quite the same as that of $\chi$ itself.  

We end this discussion with a word of caution:  Although mapping to 
an effective Gaussian is useful, it may hide interesting physics.  
For example, the non-Gaussian stochasticity in \cite{scs12} results 
in a quadrupolar signature for Lagrangian space halo bias; using 
an effective Gaussian obscures the origin of this angular dependence.  

\subsection{Excursion set peaks}\label{esp}
For walks associated with peaks in $\delta-g$, one must simply add a 
weight which depends on ${\rm d}(\delta-g)/{\rm d}s$ \citep{ms12}.  
The associated first crossing distribution becomes that for excursion 
set peaks \citep{ps12}, provided we remember to rescale
 $\delta_c(s)\to \delta_c(s)/\sqrt{1+\beta^2}$, because the peaks 
are in $g_-$ rather than in $\delta$.  Namely, 
\begin{equation}
 s f(s) = \frac{\exp(-\nu_\beta^2/2)}{2\gamma\,\sqrt{2\pi}}\,
          \int {\rm d}x\,x\,p(x|\gamma\nu_\beta)\,\frac{{\cal F}(x)}{(R_*/R)^3}
 \label{sfspk}
\end{equation}
where $\nu_\beta\equiv (\delta_c/\sigma)/\sqrt{1 + \beta^2}$ as before 
(c.f. equation~\ref{nubeta}), the parameters $\gamma$ and $R_*$ are 
defined by equation~(4.6a) in \cite{bbks86}, 
\begin{equation}
 p(x|\gamma,\nu_\beta) = \frac{{\rm e}^{-(x - \gamma\nu_\beta)^2/2(1-\gamma^2)}}
                            {\sqrt{2\pi(1-\gamma^2)}}
\end{equation}
is the usual conditional Gaussian (i.e. $\gamma$ is the correlation 
coefficent between $x$ and $\nu_\beta$), and 
${\cal F}(x)$ is given by equation~(A15) of \cite{bbks86}.  
(The Musso-Sheth approximation for the first crossing distribution for 
all walks with correlated steps has ${\cal F}(x)=1$ and $R=R_*$.)

The distribution of $\delta_{1\!\times}$ is then unchanged from that 
for all walks (equation~\ref{pd1x}), because a constraint on the 
`velocity' of $g_-$, which is what the peaks constraint boils down 
to \citep{ms12}, means nothing for $g_+$, which is what determines 
$\delta_{1\!\times}$.  
The statistics of walks centered on a randomly chosen particle 
within a protohalo are known to be different from those centered 
on the protohalo center of mass; the latter yield larger values 
of $\delta_{1\!\times}$ \citep{smt01,arsc13,dts13}.  Therefore, the 
analysis above indicates that a model which identifies protohalo 
centers of mass with peaks in $\delta-g$ cannot explain this difference.  

\begin{figure}
 \centering
 \includegraphics[width=\hsize]{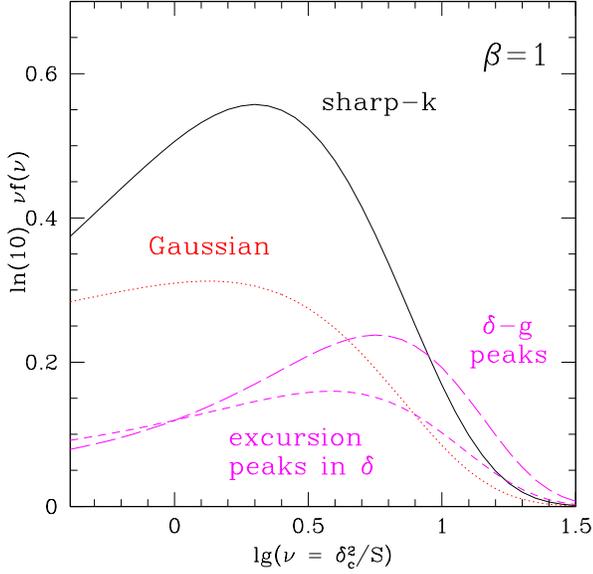}
 \caption{First crossing distribution for all walks when steps are 
          uncorrelated (solid); when steps are correlated because 
          of Gaussian smoothing and the power-spectrum is  
          $P(k)\propto k^{-1.2}$; when the walks are centered on peaks 
          in $\delta-g$ (equation~\ref{sfspk}) and on peaks in $\delta$ 
          only (equation~\ref{sfspkd}).   
         }
 \label{sfspks}
\end{figure}

If we identify protohalo centers of mass on scale $s$ with positions 
where $\delta$ first exceeds $\delta_c + g$ and are peaks in $\delta$ 
(rather than in $\delta-g$) on that scale, then the first crossing 
distribution becomes 
\begin{equation}
 s f(s) = \frac{\exp(-\nu_\beta^2/2)}{2\gamma\,(R_*/R)^3\sqrt{2\pi}}\,
  \int {\rm d}x\,x\,p(x|\gamma,\nu_\beta)\,G_0(x,\gamma_{_\beta} x)
 \label{sfspkd}
\end{equation}
where we have defined 
$\gamma^2_\beta \equiv (1 + \beta^2)^{-1}$, 
$\nu_\beta, \gamma, R_*$ and $p(x|\gamma,\nu_\beta)$
were defined above, 
and 
\begin{equation}
 G_n(x,\gamma_{_\beta} x) \equiv \int dy\, p(y|\gamma_{_\beta}, x)\, {\cal F}(y)\,y^n
 \label{G0bbks}
\end{equation}
with ${\cal F}(y)$ the same quantity that appears in 
equation~(\ref{sfspk}), i.e., given by equation~(A15) of \cite{bbks86}.
Similarly, a little algebra shows that in this case the distribution 
of $\delta_{1\!\times}$ is given by 
\begin{equation}
 p_{\rm pk}(\delta_{1\!\times}|s) =  A_{1\!\times}\,p(\delta_{1\!\times}|s)\,
   \left[G_1 - \gamma(\nu_{1\!\times}-\nu_c)\,G_0\right],
 \label{pd1xpk}
\end{equation}
where $p(\delta_{1\!\times}|s)$ is the distribution for all walks 
(equation~\ref{pd1x}), $G_n(\nu_{1\!\times},\gamma\nu_{1\!\times})$ is given 
by equation~(\ref{G0bbks}), and
 $A_{1\!\times} = [\sqrt{1+\beta ^2}
         \int {\rm d}x\,x\,p(x|\gamma,\nu_\beta)\,G_0(x,\gamma_{_\beta} x)]^{-1}$ 
is a normalization factor which ensures that the integral over all 
$\delta_{1\!\times}$ yields unity.  

\begin{figure}
 \centering
 \includegraphics[width=\hsize]{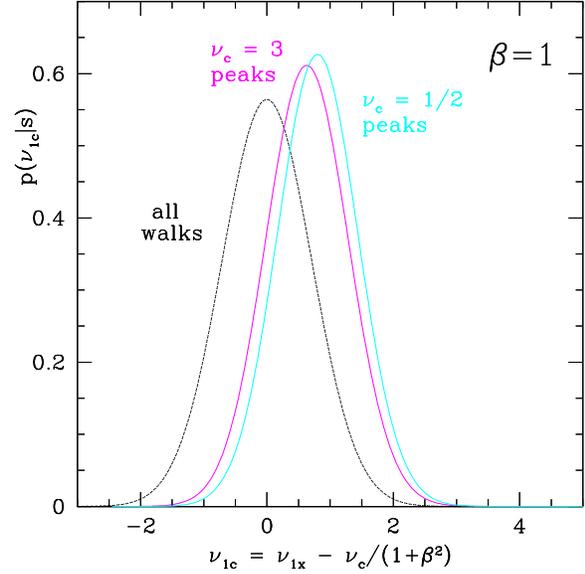}
 \caption{Distribution of walk height at first crossing, $\delta_{1\!\times}$, 
          for all walks (dotted) and for walks which are also peaks 
          in $\delta$ on scale $s$ (solid); i.e., equations~(\ref{pd1x}) 
          and~(\ref{pd1xpk}) respectively.  
          The distribution for peaks in $\delta-g$ is also given by the 
          dotted curve.  
         }
 \label{pd1xpks}
\end{figure}

In the limit $\beta\to 0$, the distribution $p(y|\gamma_{_\beta}, x)$ 
becomes sharply peaked around its mean value $\gamma_{_\beta} x\to x$, 
so that $G_0(x,\gamma_{_\beta} x)\to {\cal F}(x)$.  Thus, in this 
limit, equation~(\ref{sfspkd}) reduces to equation~(\ref{sfspk}).  
Similarly, 
 $p(\delta_{1\!\times}|s)$ becomes a delta function centered on
 $\delta_c/(1+\beta^2)$, making
 $p_{\rm pk}(\delta_{1\!\times}|s)\to A_{1\!\times}\,p(\delta_{1\!\times}|s)\,G_1$.  
Since $A_{1\!\times}\to G_1^{-1}$ in this limit, 
 $p_{\rm pk}(\delta_{1\!\times}|s)\to p(\delta_{1\!\times}|s)$ as it should.  

In general, at large $\nu_{1\!\times}$, $G_1/G_0\to \gamma\nu_{1\!\times}$ 
making 
 $p_{\rm pk}(\delta_{1\!\times}|s) \propto
  p(\delta_{1\!\times}|s)\,G_0(\nu_{1\!\times},\gamma\nu_{1\!\times})$; this 
illustrates that the term in square brackets acts to skew the 
distribution towards larger $\delta_{1\!\times}$.  
Figures~\ref{sfspks} and~\ref{pd1xpks} show this explicitly:  
they compare $sf(s)$ and $p(\delta_{1\!\times}|s)$ for these two peak 
models with that for all walks.  
In practice, we use equations~(4.4) and~(6.13) of \cite{bbks86} to 
approximate $G_0$ and $G_1/G_0$, and we assumed Gaussian smoothing of 
a scale-free power spectrum, i.e. $P(k)\propto k^n$, for which 
$\gamma^2 = (n+3)/(n+5)$ and $(R_*/R)^2 = 6/(n+5)$.  To make the 
Figures, we set $n=-1.2$ and $\beta=1$ to highlight the effects of 
$\beta$.  

Figure~\ref{sfspks} shows that peaks in $\delta$ and 
$\delta-g$ do indeed produce different counts (short and long dashed 
curves, respectively); both are different from the result for all 
walks (dotted).  And Figure~\ref{pd1xpks} shows that the distribution 
given in equation~(\ref{pd1xpk}) is indeed shifted to larger values of 
$\delta_{1\!\times}$, with the shift depending weakly on the mass 
scale $\nu_c$.  This increase in $\delta_{1\!\times}$ is qualitatively 
in the right direction, suggesting that identifying protohalos with 
peaks in $\delta$ is a better model than one where protohalos are 
identified with peaks in $\delta-g$.  However, the predicted 
distribution for peaks is not as different from that for all walks as 
is the difference seen in simulations between centre-of-mass walks and 
randomly chosen ones (the shift in the mean is not large enough, 
the width is not narrow enough, and the shape is not skewed enough).

Before moving on, we note that, in the one-dimensional problem, the 
peaks motivated approach is attractive because it provides a natural 
reason why halo counts in simulations do not fall as steeply as 
$\exp(-\delta_c^2/2s)$ at small $s$.  The two-Gaussian model here 
achieves this by setting $\beta\approx 0.6$ (see discussion at end 
of Section~\ref{universal}).  The analysis above indicates that peaks 
in this two-Gaussian model will require a smaller value of $\beta$ to 
reproduce the halo counts.  Then reproducing the distributions of 
$p(\delta_{1\!\times}|s)$ {\em and} $p_{\rm pk}(\delta_{1\!\times}|s)$ 
provide important self-consistency tests.  Since reducing $\beta$ 
from the value used to make Figure~\ref{pd1xpks} will only make all 
the curves there more similar to one another, this will exacerbate 
the discrepancies between model and simulations.  Thus, our analysis 
suggests that neither of the peaks models we have considered here are 
consistent with measurements.  

\section{Discussion}
We described a two-dimensional excursion set model, for Gaussian 
walks in $\delta$ and $g$, for which almost all quantities associated 
with first crossing distributions can be computed analytically.  
We have tested all the analytic expressions we provide in this paper 
using Monte-Carlo realizations of the two-dimensional stochastic 
process, finding excellent agreement.  Since the analytic arguments 
are sufficiently simple, we have only included a few plots showing 
this agreement.  

Our predictions include the unconditional first crossing distribution 
$f(s|z)$ (Section~\ref{universal}); 
the conditional first crossing distribution for redshift $z$,
$f(s,z|S,Z)$, by walks which are known to have first crossed one 
another on scale $S<s$ at redshift $Z<z$ (Section~\ref{f10}); 
and the conditional distribution $f(s,z|S,\Delta)$ for walks which 
are constrained to have height $\Delta$ on scale $S<s$ 
(Section~\ref{fsS}).  These are usually used to model halo abundances, 
progenitor distributions, and the environmental dependence of clustering.  
In the one-dimensional case, for appropriately chosen pairs of redshift 
and environment, the progenitor and conditional distributions are the 
same.  For higher-dimensional walks, this is no longer the case:  the 
conditional distributions generically predict more massive objects 
(Figure~\ref{evolenv} and related discussion).  

Another new feature of such higher-dimensional models is the fact that 
there is, generically, a distribution of walk heights at first crossing 
$p(\delta_{1\!\times}|s)$ (Section~\ref{d1}), and an associated 
distribution of the other variable $p(g_{1\!\times}|s)$ (Section~\ref{g1}).  
For the Gaussian walks considered here, these distributions are Gaussian, 
even when the barrier height depends on the first crossing scale $s$.  
We argued that $s$-dependence of the mean barrier height, with a 
Gaussian scatter around the mean, should provide a good approximation 
even when the walks are not Gaussian (Section~\ref{ng}).  

We also argued that, because of the variable(s) which are not $\delta$, 
halo bias in these models will generally be stochastic (sometimes 
refered to as nonlocal), and the conditional distributions will 
generically exhibit assembly bias, even when the steps in the walks 
are uncorrelated.  We provided explicit expressions for both the 
stochastic (Section~\ref{stocbias}) and the assembly bias 
(Section~\ref{abias} and Figure~\ref{cond1x}).  Although our 
model predicts no correlation between halo formation times and 
environment (at fixed halo mass), in agreement with the one-dimensional 
case, it nevertheless predicts that halos surrounded by overdensities 
should be denser and more concentrated than halos of the same mass 
in underdensities.  

The lack of correlation between time and environment is a consequence 
of studying walks with uncorrelated steps.  
We sketched how to generalize our results to include correlations 
between the steps in each walk (necessary for quantitative comparison 
with simulations; Section~\ref{cs}), and between the walks themselves 
(as might arise in models where the critical density required for 
collapse is determined by the overdensity on large scales; 
Section~\ref{cw}).  These will introduce additional assembly bias 
effects, for the same reasons they do so for one-dimensional walks.  
Although we sketched how to quantify these here, we did not show 
plots or otherwise quantify these effects for the following reason.  

One of the drawbacks of this model -- that is in common with the usual 
one-dimensional walk approach -- is that it is explicitly about the 
statistics of all points in space.  However, halos form around special 
positions in space, and the statistics of this point process -- arguably 
the point process for which the description of the physics is simplest -- 
is very different from that around randomly chosen positions 
\citep{smt01,ps12,arsc13}.  We argued that that the simplest case, in 
which halos form around positions which are peaks in $\delta-g$, cannot 
explain this difference (Section~\ref{esp}).  Although a model in which 
halos form around peaks in $\delta$ fares better (Figures~\ref{sfspks} 
and~\ref{pd1xpks}), it fails to adequately model the differences 
between walks centred on all particles, and those centred on the 
special subset which are protohalo centers of mass.  
Work in progress shows how to extend this approach to include a more 
elaborate model for protohalo centers-of-mass, but we believe our 
results demonstrate the power of requiring models to reproduce not 
just halo counts but the distribution of $\delta$ at fixed halo mass 
as well.

\section*{Acknowledgements}
This work is supported in part by NSF-0908241 and NASA NNX11A125G.
RKS thanks the LUTH group at Meudon Observatory for hospitality
during the summer of 2012, I. Achitouv for discussions about 
the conditional crossing distribution, and A. Paranjape for 
discussions about excursion set peaks.

\appendix


\section{Proof of equation~(15)}
The main complication with respect to the one dimensional case is 
that the constraint that the walk passed through $\Delta$ on scale 
$S$ still allows walks with a range of values of $G$.  This range 
is constrained by the requirement that $\Delta$ and $G$ had not 
crossed on scales smaller than $S$.  At fixed $\Delta$ {\em and} 
$G$, the solution is straightforward, as we show shortly, so the 
main work is to integrate this solution over the allowed range of $G$.  

As before, it is best to work in the $(g_{+},g_{-})$ plane, in which 
case the requirement that the walk has height $\Delta$ on scale $S$ 
means that 
\begin{equation}
 G_{-} < \frac {\delta_c(S)} {\sqrt{1+\beta^2}} \quad {\rm and} \quad
 G_{+} = \Delta \sqrt{1+\beta^2} - G_{-} 
\end{equation}
(again capital letters indicate values at $S$).  The distribution 
of $s$ at which $\delta_c(s)/\sqrt{1+\beta^2}$ is first crossed, 
given that the walk started from $(G_+,G_-)$ on scale $S$, is given 
by equation~(\ref{sfs}) with 
\begin{equation}
 \nu^2 = \frac {(\delta_c/ \sqrt{1+\beta^2} - G_{-})^2} {s-S}.
\end{equation}
Notice that this expression depends only on $G_-$, so we will 
denote the associated first crossing distribution as $f(s|G_-,S)$.  

To get the quantity we are after, $f(s|\Delta,S)$, we must now 
integrate $f(s|G_-,S)$ over all allowed starting values $(G_+,G_-)$, 
weighting by the probability of starting at each.  I.e., 
\begin{align}
 f(s|\Delta,S) &= A \int^{\infty}_{-\infty} {\rm d}G_{+} \; 
               \int^{\delta_{c}/\sqrt{1+\beta^2}}_{-\infty}  {\rm d}G_{-} \,
              f(s|G_{-},S)\nonumber \\
 &\qquad \times \quad p(G_{+}|S) \, q( G_{-} | S) \,\delta_{\rm D}(\delta-\Delta)
 \label{cond_fs}
\end{align}
where
\begin{equation}
 q(G_{-}|S)= \frac{{\rm e}^{-G_{-}^2/2S}}{\sqrt{2\pi S}}
      - \frac{{\rm e}^{-(2\delta_c / \sqrt{1+\beta^2}-G_{-})^2/2S}}{\sqrt{2\pi S}}
\end{equation}
is the probability that (the one-dimensional) walk $g_-$ has height $G_{-}$ 
at $S$ and never crossed $\delta_c/\sqrt{1+\beta^2}$ on some smaller $s<S$ 
\citep{bcek91}, 
$p(G_{+}|S)$ is a Gaussian with zero mean and variance $S$, 
and 
\begin{equation}
 A \equiv \int^{\infty}_{-\infty} {\rm d}G_{+} 
  \int^{\delta_{c}/\sqrt{1+\beta^2}}_{-\infty} \!\!\! {\rm d}G_{-}\,
   p(G_{+}|S) \, q(G_{-}|S) \,\delta_{\rm D}(\delta-\Delta) 
\end{equation}
is a normalization constant which ensures that the probabilities 
integrate to unity.  
This, and the integral in eq.(\ref{cond_fs}) can be performed 
analytically, yielding 
\begin{equation}
 f(s|\Delta,S) = \frac{(\delta_c-\Delta)(1+\beta^2)}{s-S + \beta^2 s}
                 \frac{{\rm exp}^{-(\delta_c - \Delta)^2/2(s-S + \beta^2 s)}}
                  {\sqrt{2\pi(s-S + \beta^2 s)}}.
\end{equation}
This is equivalent to the change of variables given by 
equation~(\ref{nuD}) of the main text.

\label{lastpage}

\end{document}